\documentclass[12pt]{article}
\usepackage{pic02}
\usepackage{hyperref}
\usepackage{url}
\usepackage{graphicx}

\begin{document}


\title{\bf ATMOSPHERIC AND LONG BASELINE NEUTRINO}
\author{T. Nakaya \\
{\em Kyoto University, Faculty of Science, Kyoto 606-8502, JAPAN}\\
for Super-Kamiokande and K2K Collaborations
}
\maketitle

%
%
%
%
%
%
\vspace{4.5cm}
%

\baselineskip=14.5pt
\begin{abstract}
 This paper presents the recent results from Super-Kamiokande atmospheric 
neutrinos and from K2K accelerator neutrinos.
Both results show the signal of neutrino oscillation, and 
provide new and precise information of oscillation parameters. 
The oscillation parameters are constraint to be
between $1.5\times 10^{-3}$~$\rm  eV^2$ and $3.9\times 10^{-3}$~$\rm eV^2$ for
$\Delta m^2$ and to be greater than 0.92 for 
${\rm sin^22\theta}$.

\end{abstract}
\newpage

\baselineskip=17pt

\section{Introduction}
After discovery of neutrino oscillation in atmospheric neutrinos by
Super-Kamiokande (SK)~\cite{sk}, much attentions are attracted by this new 
phenomena. In this paper, the new results from SK atmospheric neutrinos
and the new result from K2K accelerator long baseline neutrino experiment
are presented. The paper only provides the digest of new results, and
does not provide the description of the experiments and the analysis.
The description of the experiments is found in references~\cite{sk,k2k} and the
description of the analysis is found in NEUTRINO 2002 conference~\cite{nu02}.

\section{Atmospheric Neutrino}
The recent results on atmospheric neutrinos from Super-Kamiokande (SK)
experiment are presented in this paper.
The observation of atmospheric neutrinos by SK is based on 1,489 
live-day exposure.
The standard selection criteria and analysis techniques are shown in 
the reference~\cite{sk}, and are not explained in this paper.

At first, the number of contained events with the vertex inside of the 
fiducial volume of the detector are shown in Table~\ref{atmpd}.
\begin{table}[hbp]
\begin{center}
\caption{ \it The number of atmospheric neutrino events and the MC expectation.}
\vskip 0.1 in
\begin{tabular}{|l|c|c|} \hline
Event category & \# DATA & \# MC (Honda)\\
\hline
\hline
Sub-GeV FC 1 ring $e-like$   event   & 3266  & 3081.0  \\
Sub-GeV FC 1 ring $\mu-like$ event   & 3181  & 4703.9  \\
Sub-GeV FC multi-ring event          & 2457  & 2985.6  \\
\hline
Sub-GeV FC Total                      & 8904  & 10770.5 \\
\hline \hline
Multi-GeV FC 1 ring $e-like$   event   & 772   & 707.8 \\
Multi-GeV FC 1 ring $\mu-like$ event   & 664   & 968.2  \\
Multi-GeV FC multi-ring event          & 1532  & 1903.5  \\
\hline
Multi-GeV FC Total                      & 2968  & 3579.4 \\
\hline \hline
PC event & 913 & 1230.0 \\
\hline
\end{tabular}
\label{atmpd}
\end{center}
\begin{footnotesize}
\hspace*{0.7cm}{\bf Sub-GeV :} Deposited Energy in SK is less than 1.33~GeV.\\
\hspace*{0.7cm}{\bf Multi-GeV :} Deposited Energy in SK is larger than 1.33~GeV.\\
\hspace*{0.7cm}{\bf FC :} Fully Contained events with both the vertex and the stopping point inside of SK.\\
\hspace*{0.7cm}{\bf PC :} Partially Contained events with the vertex inside and the stopping point outside of SK.\\
\hspace*{0.7cm}{\bf ring :} The number of ``ring'' corresponds to the number of observed particles.\\
\end{footnotesize}
\end{table}
The observed number of muon neutrinos are smaller than the Monte Carlo (MC) 
prediction. The double ratio of the number of $\mu-like$ events and the number of
$e-like$ events between data and MC are:\\
\hspace*{1.2cm}
\begin{math}
\frac{(\mu/e)_{DATA}}{(\mu/e)_{MC}} = 0.638^{+0.016}_{-0.016} \pm 0.050 \mbox{ (Sub-GeV FC)} 
\end{math}
\\
\hspace*{1.2cm}
\begin{math}
\frac{(\mu/e)_{DATA}}{(\mu/e)_{MC}} = 0.658^{+0.030}_{-0.028} \pm 0.078 \mbox{ (Multi-GeV FC)} 
\end{math}
\\ The flux of upward through-going muon originated by higher energy 
neutrinos is
$1.7 \pm 0.04 \pm 0.02$~$ (\times 10^{-13} cm^{-2} s^{-1} sr^{-1})$  with the 
expectation of $1.97 \pm 0.44$~$ (\times 10^{-13} cm^{-2} s^{-1} sr^{-1})$.
The flux of upward stopping muon is 
$0.41 \pm 0.02 \pm 0.02$~$ (\times 10^{-13} cm^{-2} s^{-1} sr^{-1})$  with the 
expectation of $0.73 \pm 0.16$~$ (\times 10^{-13} cm^{-2} s^{-1} sr^{-1})$.
All measurements show the deficit of muon neutrinos.

\begin{figure}[hbtp]
\begin{center}
\includegraphics[width=14cm]{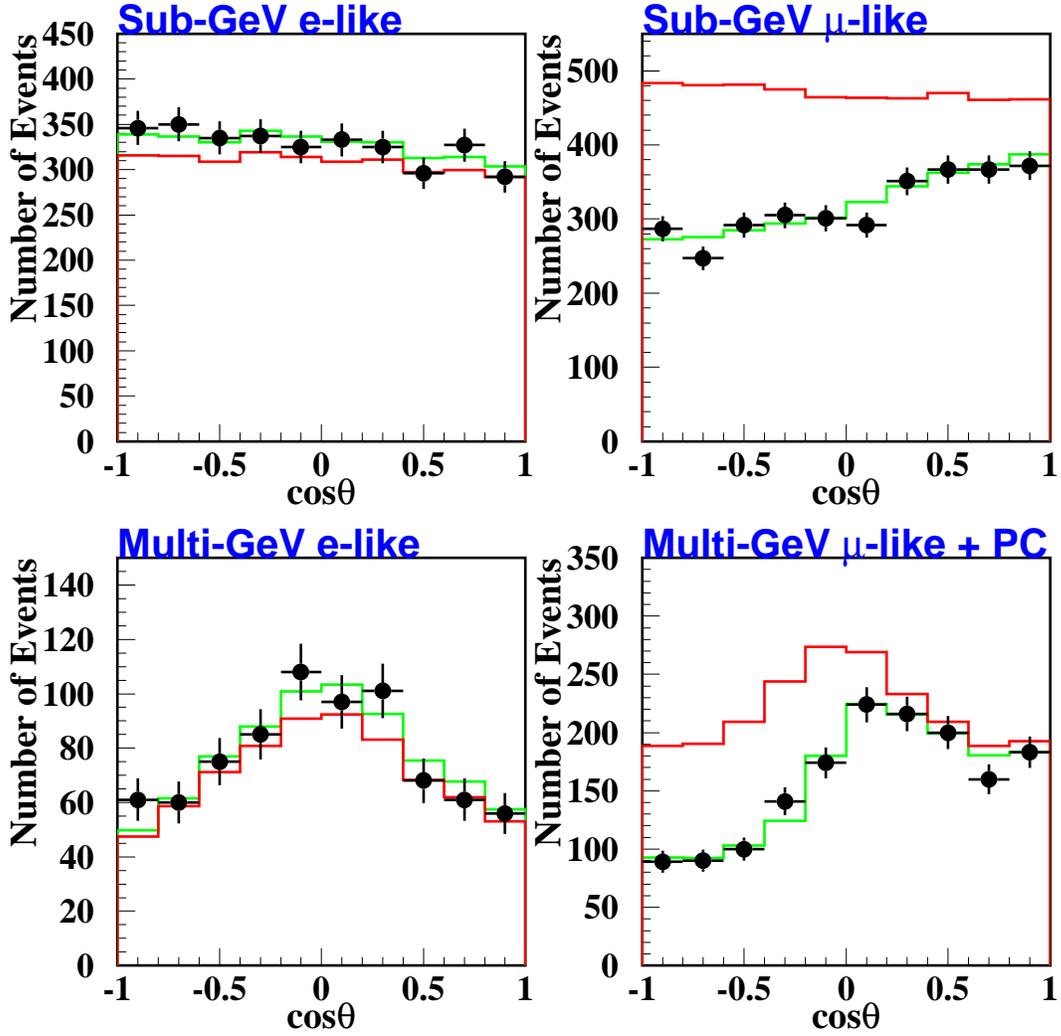}
 \caption{\it
Zenith angle distribution of Sub-GeV
single ring e-like events, $\mu$-like events, Multi-GeV single ring e-like
events and $\mu$-like events + P.C. events.
Closed circle, red histogram and green histogram are data, atmospheric neutrino 
MC events w/o neutrino oscillation and best fit oscillated expectation
with ${\rm \Delta m^2=2.5\times 10^{-3}eV^2}$ and ${\rm sin^22\theta=1.0}$
    \label{sk1} }
\end{center}
\end{figure}

\begin{figure}[hbtp]
\begin{center}
\includegraphics[width=14cm]{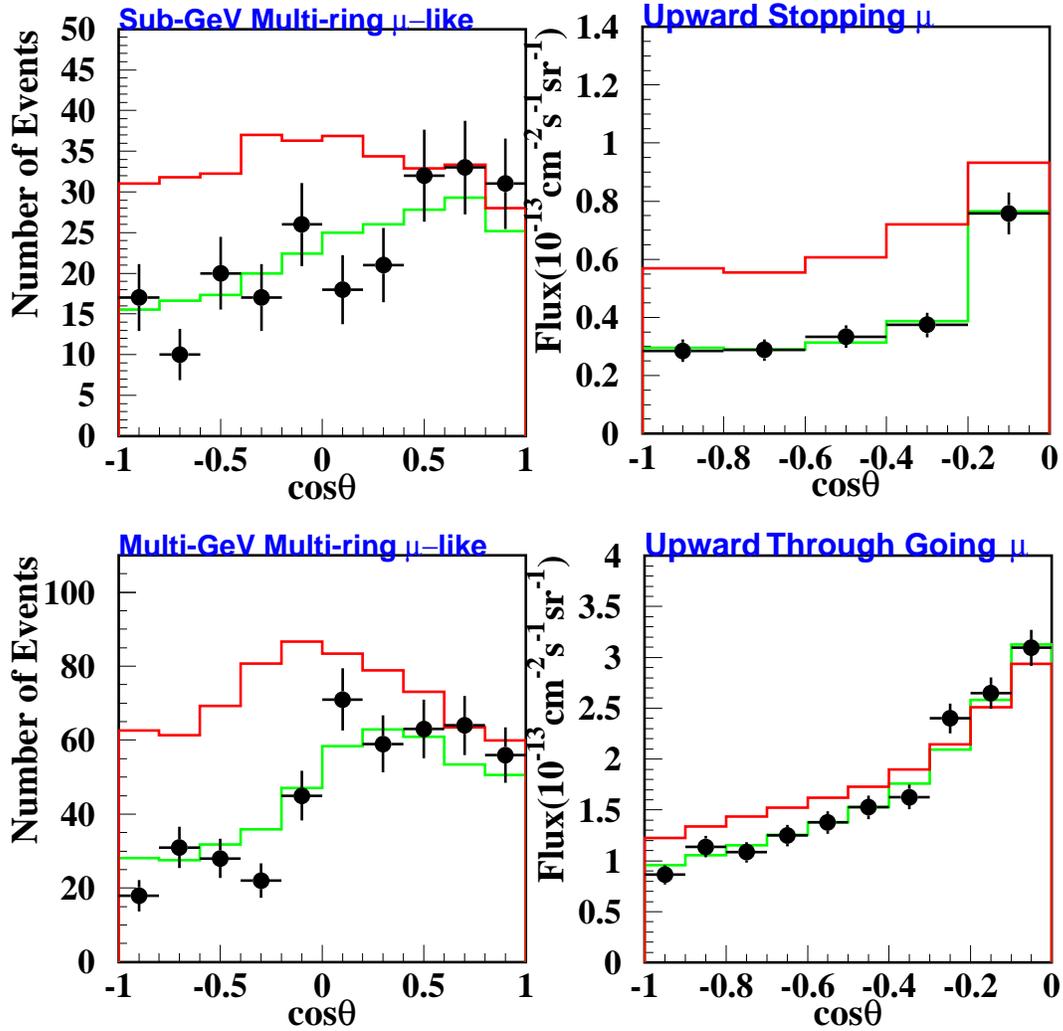}
 \caption{\it
Zenith angle distribution of multi-ring $\mu$-enrich data,
upward through going $\mu$ and upward stopping $\mu$.
Closed circle, red histogram and green histogram are data, atmospheric neutrino 
MC events w/o neutrino oscillation and best fit oscillated expectation
with ${\rm \Delta m^2=2.5\times 10^{-3}eV^2}$ and ${\rm sin^22\theta=1.0}$
    \label{sk2} }
\end{center}
\end{figure}

\begin{figure}[hbtp]
\begin{center}
\includegraphics[width=10cm]{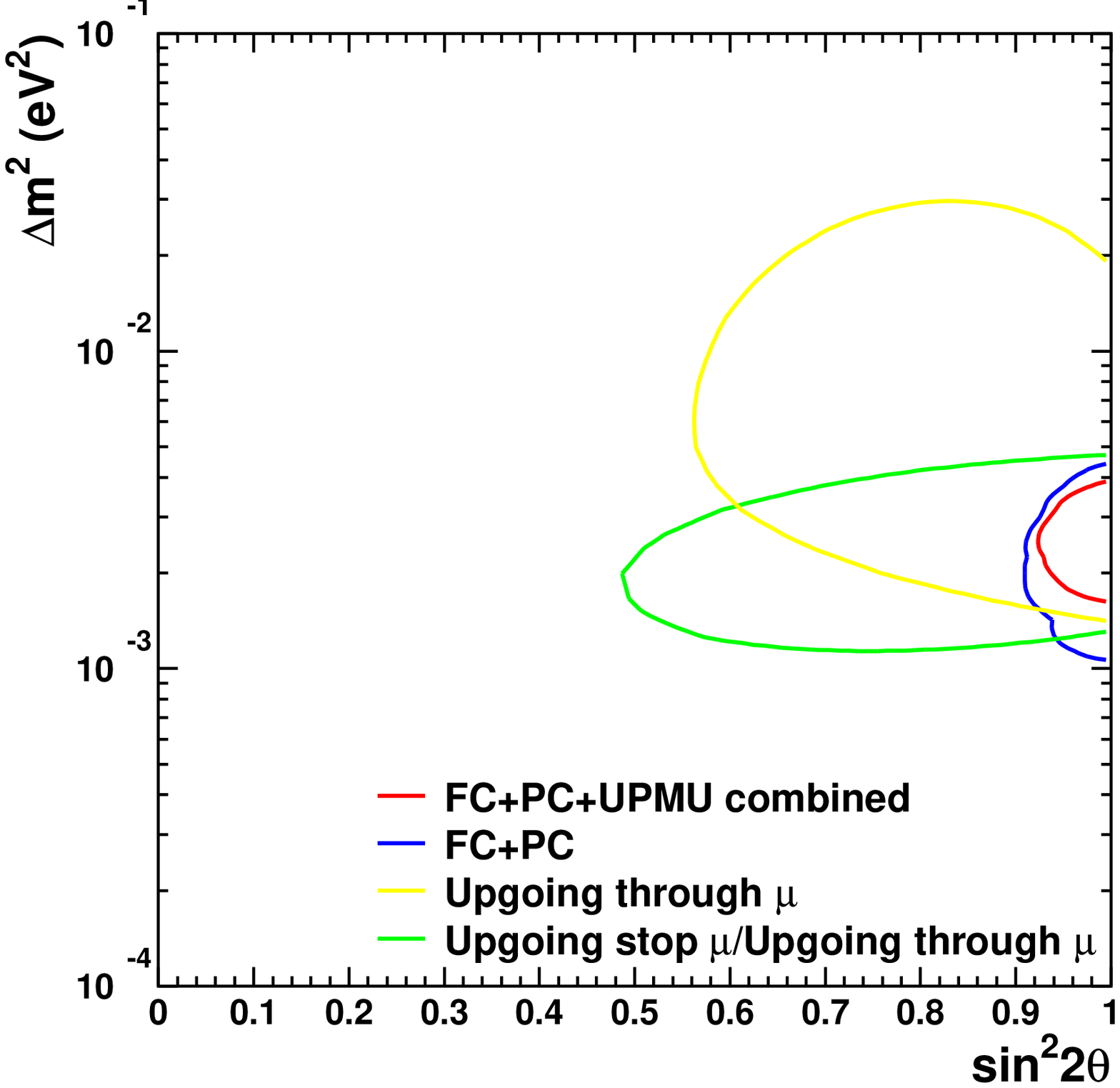}
 \caption{\it
The 68\%, 90\% and 99\% C.L. allowed region using all atmospheric neutrino 
events (F.C. 1-ring, F.C. multi-ring, P.C., upward through going $\mu$ and 
upward stopping $\mu$) for $\nu_\mu-\nu_\tau$ 2 flavor neutrino oscillation
in Super-Kamiokande.
    \label{sk3} }
\end{center}
\end{figure}

The zenith angle distributions are shown in Figure~\ref{sk1} and
\ref{sk2}.
The distributions are well explained by neutrino oscillation.
The best fit point is 
${\rm \Delta m^2=2.5\times 10^{-3}eV^2}$ and ${\rm sin^22\theta=1.0}$
with $\chi^2_{min}. = 163.2/170$~$ dof$ ($dof$: degree of freedom).
In the case of no oscillation, the $\chi^2_{min}$ is $456.5/172$~$dof$.
Null oscillation scenario is completely ruled out.
The contour plot of neutrino oscillation is shown in Figure~\ref{sk3}.
At 90\% C.L., the allowed region of $\Delta m^2$ is in the range of 
$(1.6-3.9) \times 10^{-3}$~$\rm eV^2$, and $\sin^2 2\theta$ is greater than 0.92.
The SK oscillation analysis is robust to the uncertainty of the
atmospheric neutrino flux calculations and the neutrino interaction
cross sections.

\begin{figure}[hbtp]
\begin{center}
\includegraphics[width=6.5cm]{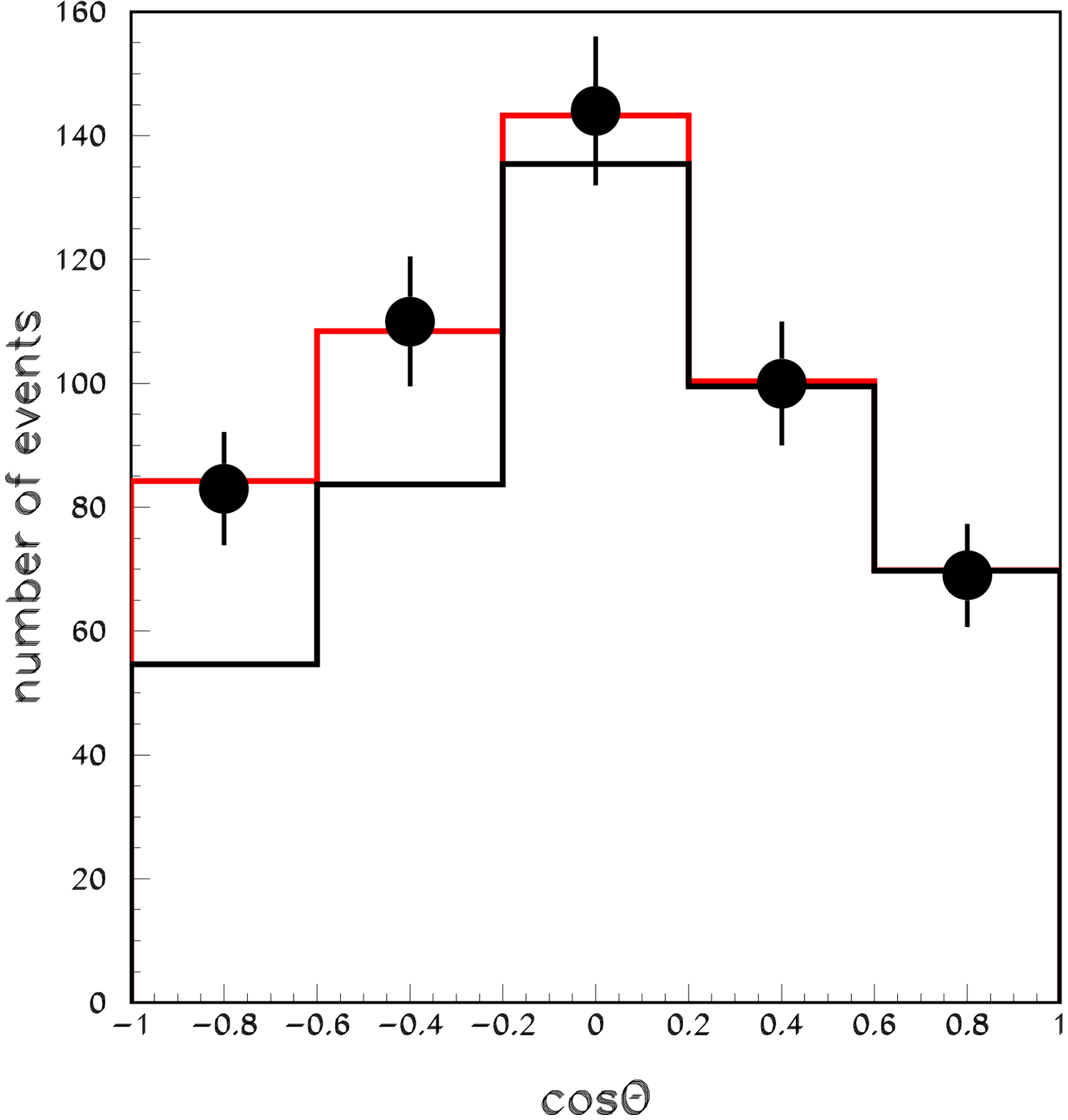}
 \caption{\it
Zenith angle distribution of tau enriched sample.
The red histogram is expectation of neutrino oscillation
with $\nu_\mu \to \nu_\tau$ channel. The black histogram is
without $\nu_\mu \to \nu_\tau$ channel.
    \label{sk4} }
\end{center}
\end{figure}

\begin{figure}[hbtp]
\begin{center}
\includegraphics[width=9.5cm]{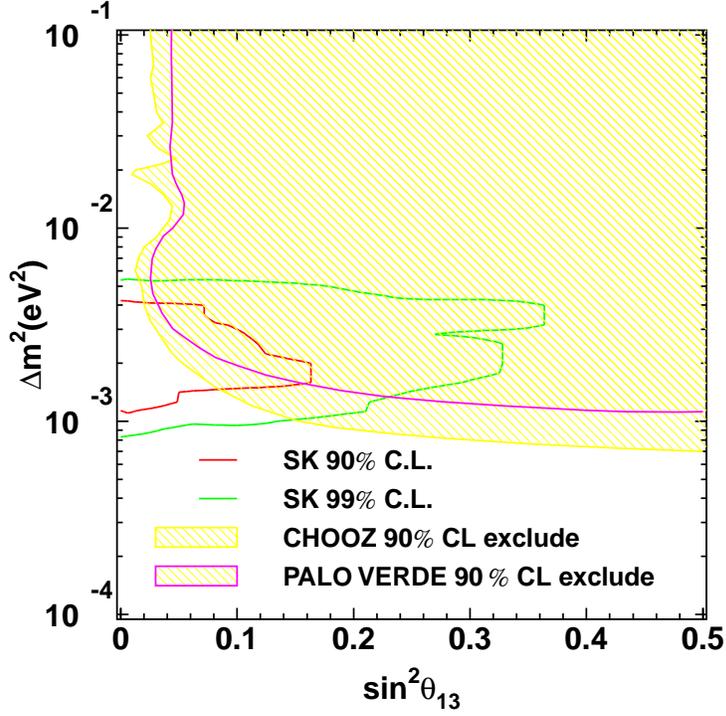}
 \caption{\it
The 90\% and 99\% C.L. allowed region in three flavor oscillation
analysis in Super-Kamiokande. 
As a reference, the results from CHOOZ and PALOVERDE experiments
are overlaid. 
    \label{sk5} }
\end{center}
\end{figure}

Since most favored channel of neutrino oscillation is $\nu_\mu \to \nu_\tau$,
the $\tau$ analysis is performed to enhance the fraction of $\tau$ events
in atmospheric neutrino data.
The number of $\tau$ neutrino candidates in FC sample is
$N^{FC}_\tau = 145 \pm 44 (stat.) ^{+11}_{-16} (sys.)$ with
the MC expectation of 86.
The zenith angle distribution of $\tau$ candidate is shown in Figure~\ref{sk4}.
The distribution is consistent with $\nu_\mu \to \nu_\tau$ oscillation.
The three flavor oscillation analysis is performed, and limit
on $\theta_{13}$ from $\nu_\mu \to \nu_e$ channel.
The result shows no $\nu_\mu \to \nu_e$ oscillation and it is consistent with 
CHOOZ and PALOVERDE results.
The contour plot is shown in Figure~\ref{sk5}.

\begin{figure}[hbtp]
\begin{center}
\includegraphics[width=9.5cm]{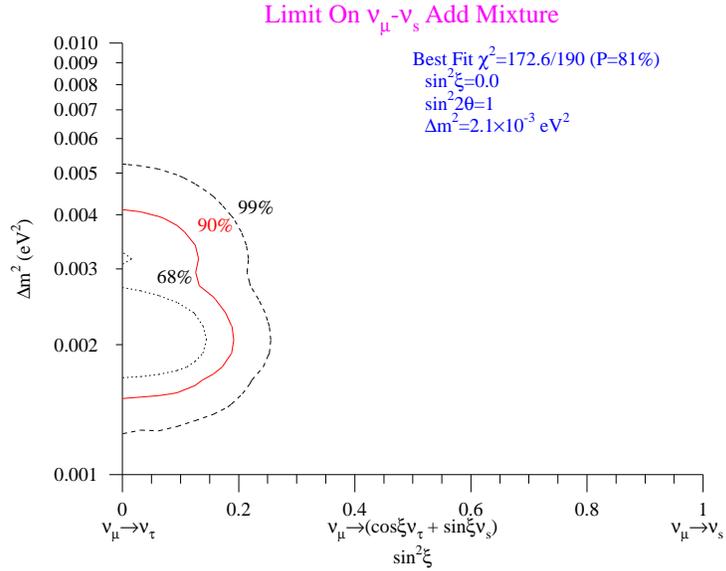}
 \caption{\it
The 68\%, 90\% and 99\% C.L. allowed region of the fraction of 
the $\nu_\mu \to \nu_\tau$ oscillation channel in 
$\nu_\mu \to \nu_\tau + \nu_{sterile}$ oscillation.
    \label{sk6} }
\end{center}
\end{figure}

Finally, the hypothesis of $\nu_\mu \to \nu_{sterile}$ oscillation
channel is tested. The fraction of $\nu_\mu \to \nu_\tau$ probability
is defined as: 
\begin{math}
\nu_{\mu} \to cos \xi \cdot \nu_\tau + sin \xi \cdot \nu_{sterile}.
\end{math}
\\ The limit on $\nu_\mu \to \nu_{sterile}$ is $sin^2 \xi < 0.19$ at 90\% C.L.,
and the result is shown in Figure~\ref{sk6}.

\newpage

\section{Long Baseline Neutrino}
The recent results from K2K, the first accelerator
long-baseline neutrino oscillation experiment from KEK to Kamioka, 
are presented in this paper. 
The K2K experiment studies neutrino oscillation 
discovered by atmospheric neutrinos~\cite{sk}.
The details of K2K experiment is described in reference~\cite{k2k}.
The neutrino oscillation probability in two flavors is expressed as
$P(\nu_\mu \to \nu_\tau) = \sin^2 2\theta \cdot 
\sin^2(\frac{1.27 \Delta m^2 L}{E_\nu})$.
As a signal of neutrino oscillation in K2K, 
both the reduction and the energy distortion of muon neutrinos at SK are 
studied. 
The data set for this analysis is based on $4.8 \times 10^{19}$ protons on 
the target delivered by KEK 12~GeV Proton Synchrotron.
 
\begin{figure}[hbtp]
\begin{center}
\includegraphics[width=6cm]{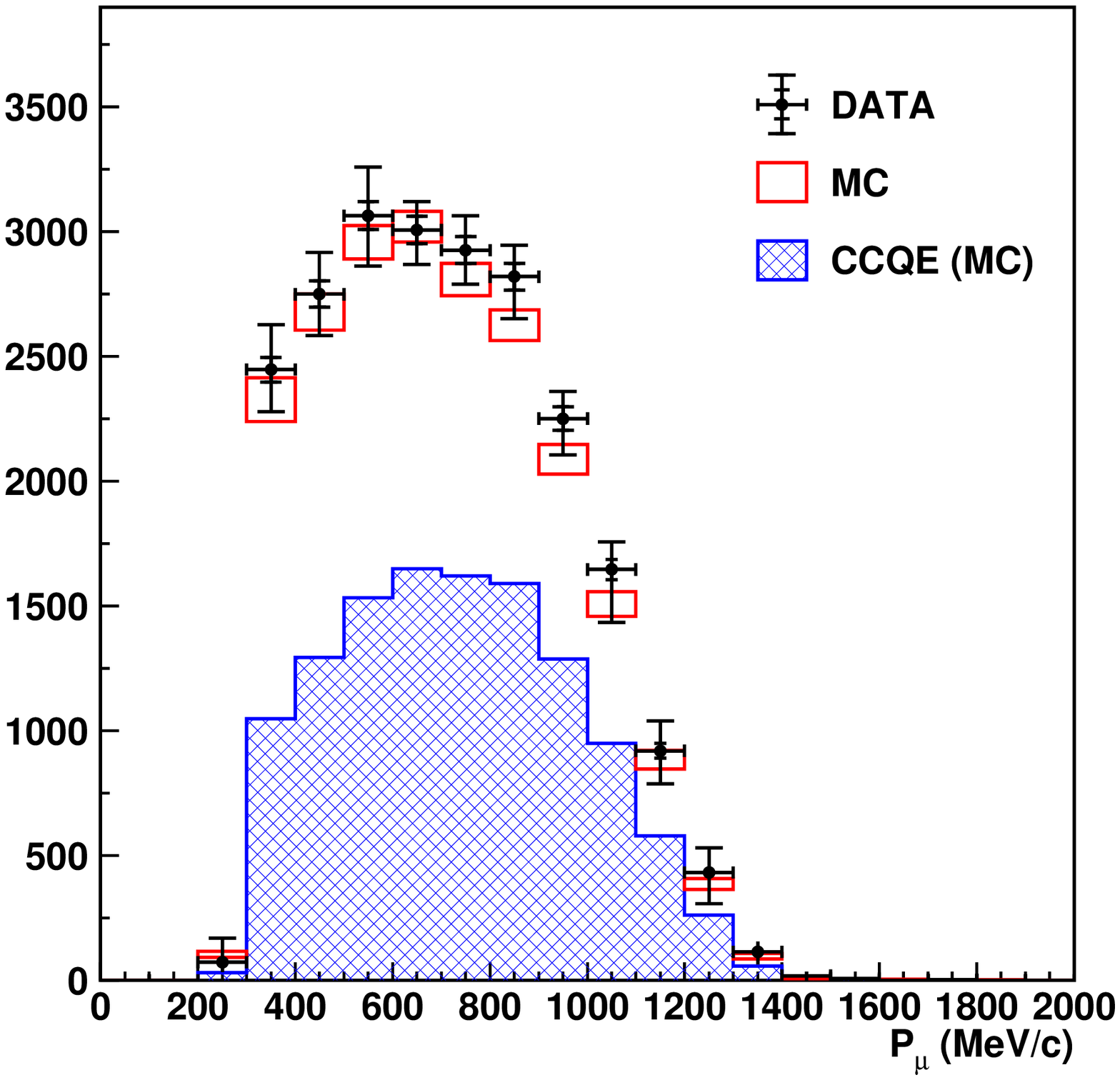}
\includegraphics[width=6cm]{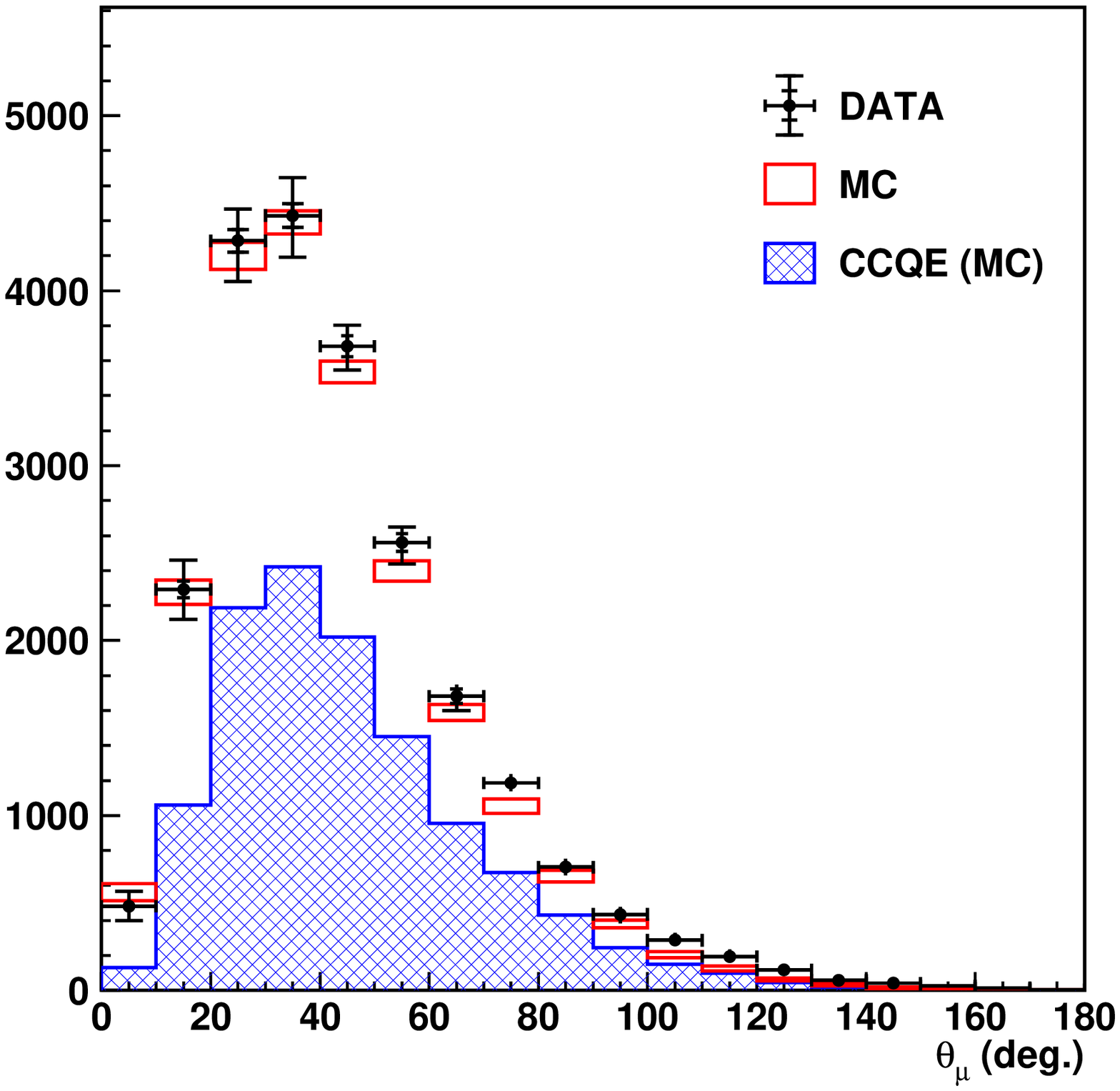}
 \caption{\it
The muon momentum and the angular distributions by K2K 1kt detector.
The crosses are data and the boxes are MC simulation with the
measured flux. The shaded histogram is the QE contribution in MC
events.
    \label{k2k1} }
\end{center}
\end{figure}

\begin{figure}[hbtp]
\begin{center}
\includegraphics[width=8cm]{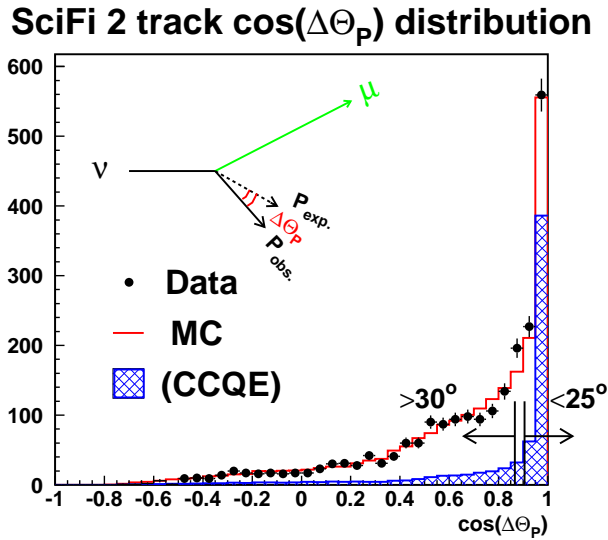}
 \caption{\it
The $\cos(\Delta\theta)$ distribution for two-track events
reconstructed in SciFi detector.
    \label{k2k2} }
\end{center}
\end{figure}

\begin{figure}[hbtp]
\begin{center}
\includegraphics[width=5.5cm]{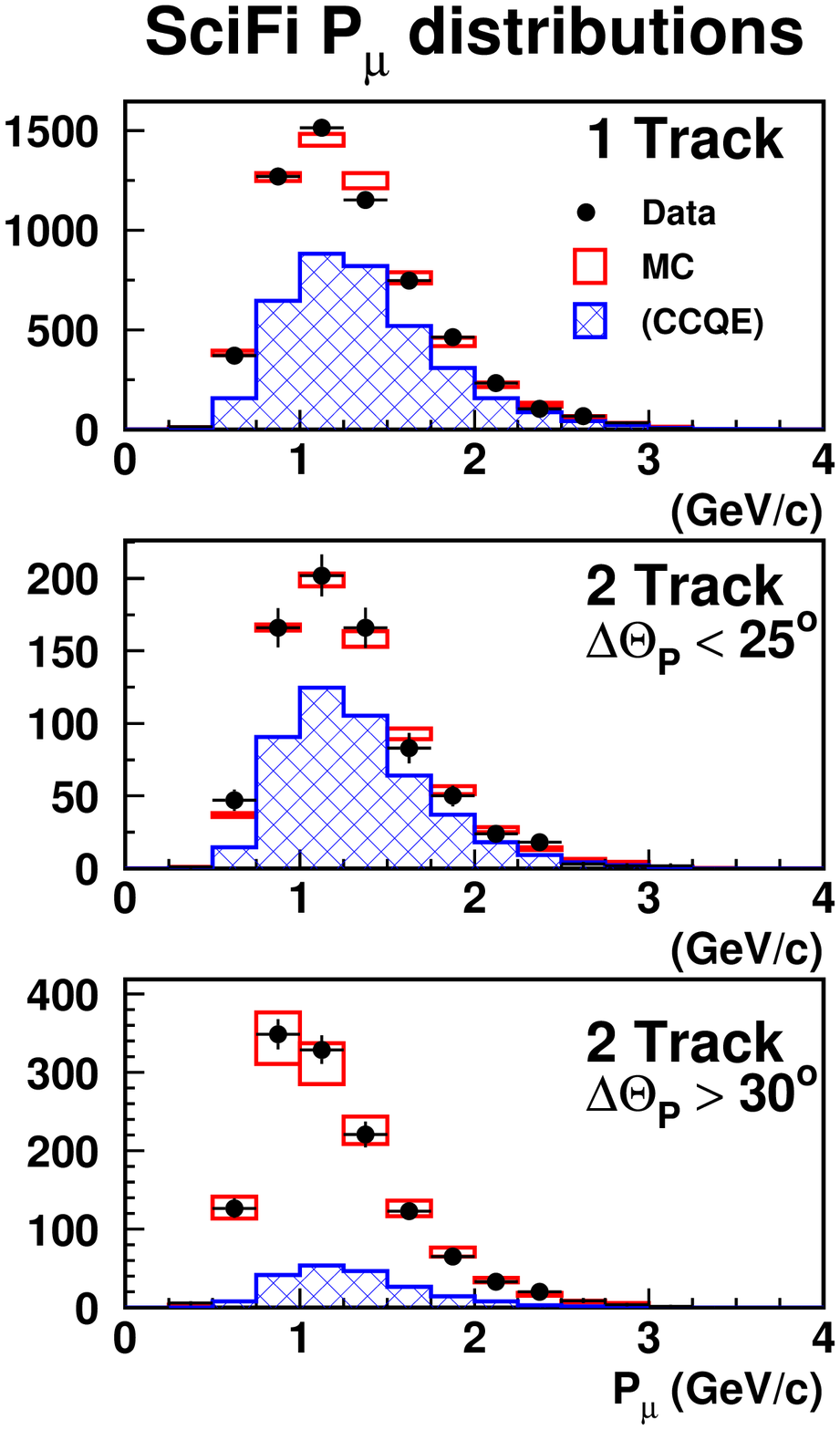}
\includegraphics[width=5.5cm]{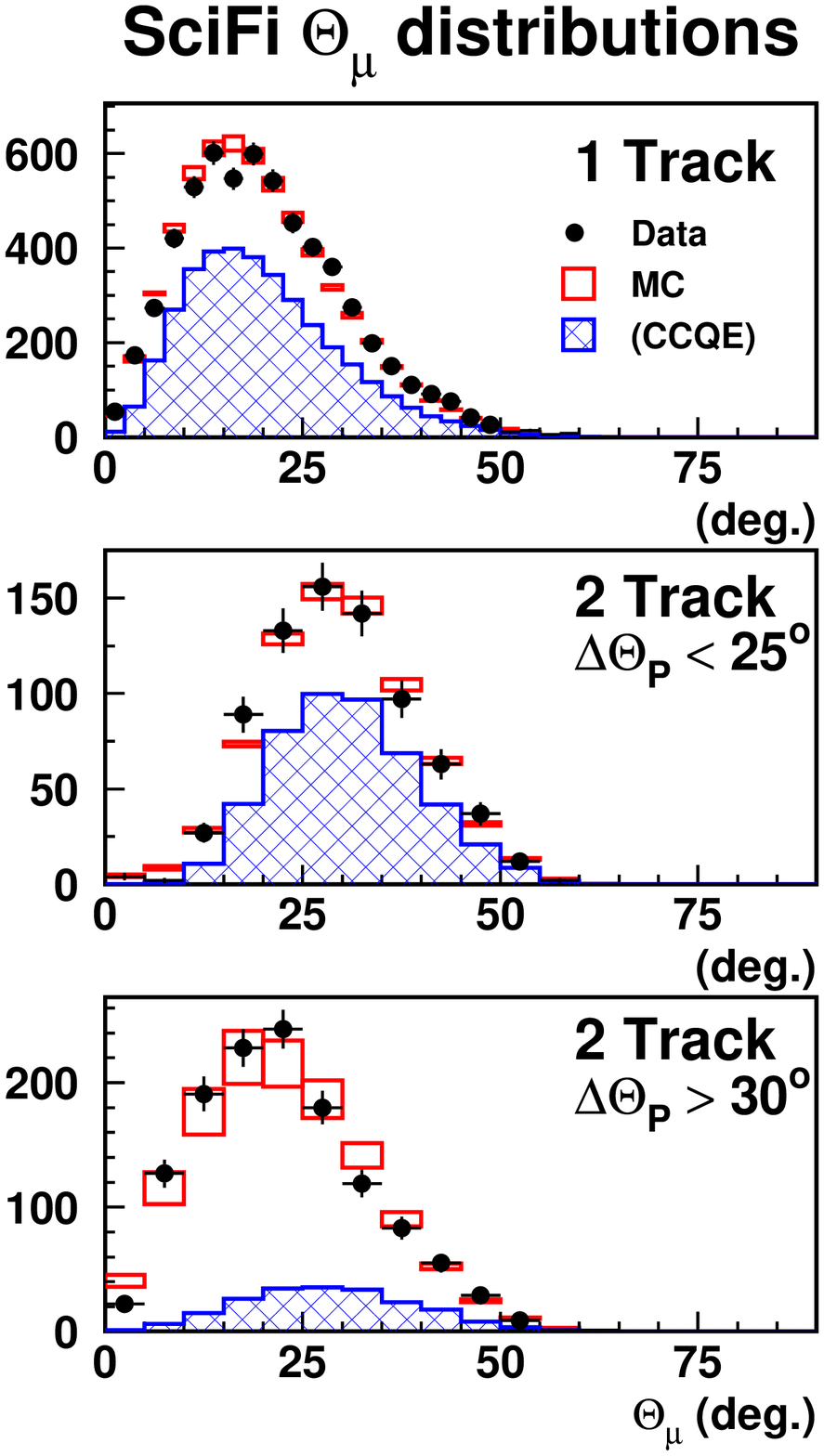}
 \caption{\it
The muon momentum and the angular distributions by K2K SciFi detector.
The crosses are data and the boxes are MC simulation with the
measured flux. The shaded histogram is the QE contribution in MC
events. The SciFi events are divided into three samples: 1-track
sample (top), 2-track QE-enriched sample (middle), and
2-track nonQE-enriched sample.
    \label{k2k3} }
\end{center}
\end{figure}

At first, the neutrino flux at KEK is measured by K2K near detectors.
The K2K near detectors used for this analysis are 
1kt water cherenkov detector (1kt) and
the scintillating fiber and water target tracker (SciFi).
The muon momentum and the angular distributions measured by
1kt are shown in Figure~\ref{k2k1}.
Single ring $\mu-like$ events are selected by 1kt for this
spectrum measurement.
%
In this analysis, the knowledge of neutrino interaction is 
important since the neutrino energy is reconstructed with an
assumption of charged current quasi-elastic (QE) interaction.
The reconstructed neutrino energy is defined as:\\
\begin{equation}
{\rm E_\nu^{rec}}=\frac {m_N E_\mu-m^2_\mu/2} {m_N-E_\mu+P_\mu cos\theta_\mu} \label{eq}, 
\end{equation}
where $\rm m_N$, $\rm E_\mu$, $\rm m_\mu$, $\rm P_\mu$ and $\theta_\mu$ are 
the nucleon mass, muon energy, mass, momentum and scattering angle 
respectively. We define nonQE interaction as all neutrino interactions 
excluding the QE interaction.
By K2K SciFi detector, the ratio of nonQE interaction to QE interaction
is measured as follows. The direction of a proton track is 
predicted from a muon track with an assumption of QE interaction.
We define the angle $\Delta\theta$ as the difference between the
prediction and the direction of an observed track.
The $\cos(\Delta\theta)$ is shown in Figure~\ref{k2k2}, in which
QE events have a peak at one. 
%
By using $\cos(\Delta\theta)$, the SciFi muon events are
divided into three categories: one is one-track sample, the second is
two-track QE-enriched sample with $\Delta\theta < 25 ^\circ$, and the third is 
two-track nonQE-enriched sample with $\Delta\theta > 30 ^\circ$.
The muon momentum and angular distributions of each sample are shown in 
Figure~\ref{k2k3}.
%
%
\begin{figure}[hbtp]
\begin{center}
\includegraphics[width=6.5cm]{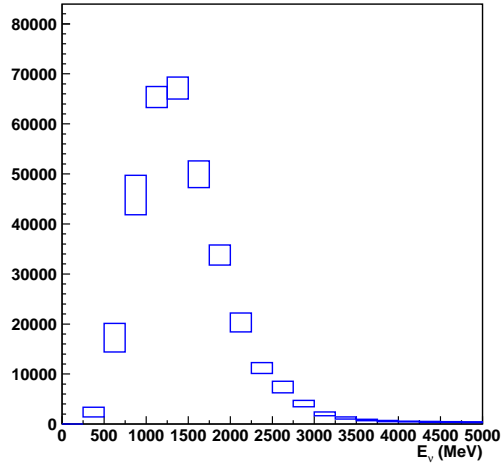}
 \caption{\it
The measured neutrino energy spectrum at KEK by K2K near detectors.
The absolute flux (vertical scale) is arbitrary.
    \label{k2k4} }
\end{center}
\end{figure}
The muon momentum and angular distributions of both 1kt and SciFi are
used to measure neutrino energy spectrum at KEK. The measured spectrum 
is shown in Figure~\ref{k2k4}. The integrated flux for normalization 
at SK is measured by 1kt.

\begin{figure}[hbtp]
\begin{center}
\includegraphics[width=10cm]{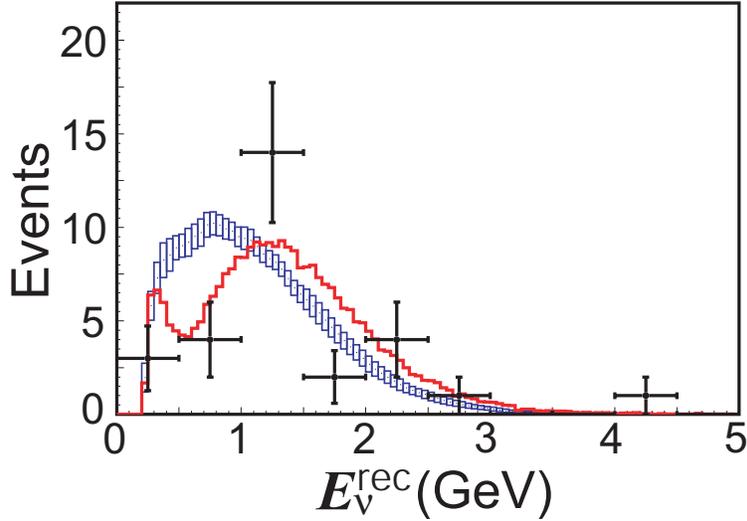}
 \caption{\it
The reconstructed $E_\nu^{rec}$ distribution by SK.
The dot is data, the box is the MC expectation without neutrino 
oscillation. The size of the box show the error of the estimation.   
The red histogram is the expectation of oscillation with the best fit parameters; 
$\rm \Delta m^2=2.8\times 10^{-3}$~$\rm eV^2$ and ${\rm sin^22\theta=1.0}$
\label{k2k5} }
\end{center}
\end{figure}

The neutrino energy spectrum at SK is estimated based on the measured spectrum
at KEK multiplying the far/near spectrum ratio estimated by the beam Monte Carlo
simulation. The description of the far/near ratio is found in the 
reference~\cite{k2k}.
We observe 56 events at SK in the fiducial volume with an estimate of 
$80.1^{+5.4}_{-6.2}$ without neutrino oscillation.
Twenty-nine single ring muon events in the fiducial volume is used for
the spectrum shape analysis.
The observed $E_\nu^{rec}$ distribution is shown in Figure~\ref{k2k5}.
The MC predictions of both with and without neutrino oscillation are 
overlaid in Figure~\ref{k2k5}. The observed $E_\nu^{rec}$ distribution
is better matched to the expected spectrum with neutrino oscillation.
The KS rest probability is 79\% for spectrum of oscillation.
%
%
%
%

\begin{figure}[hbtp]
\begin{center}

\includegraphics[width=6cm]{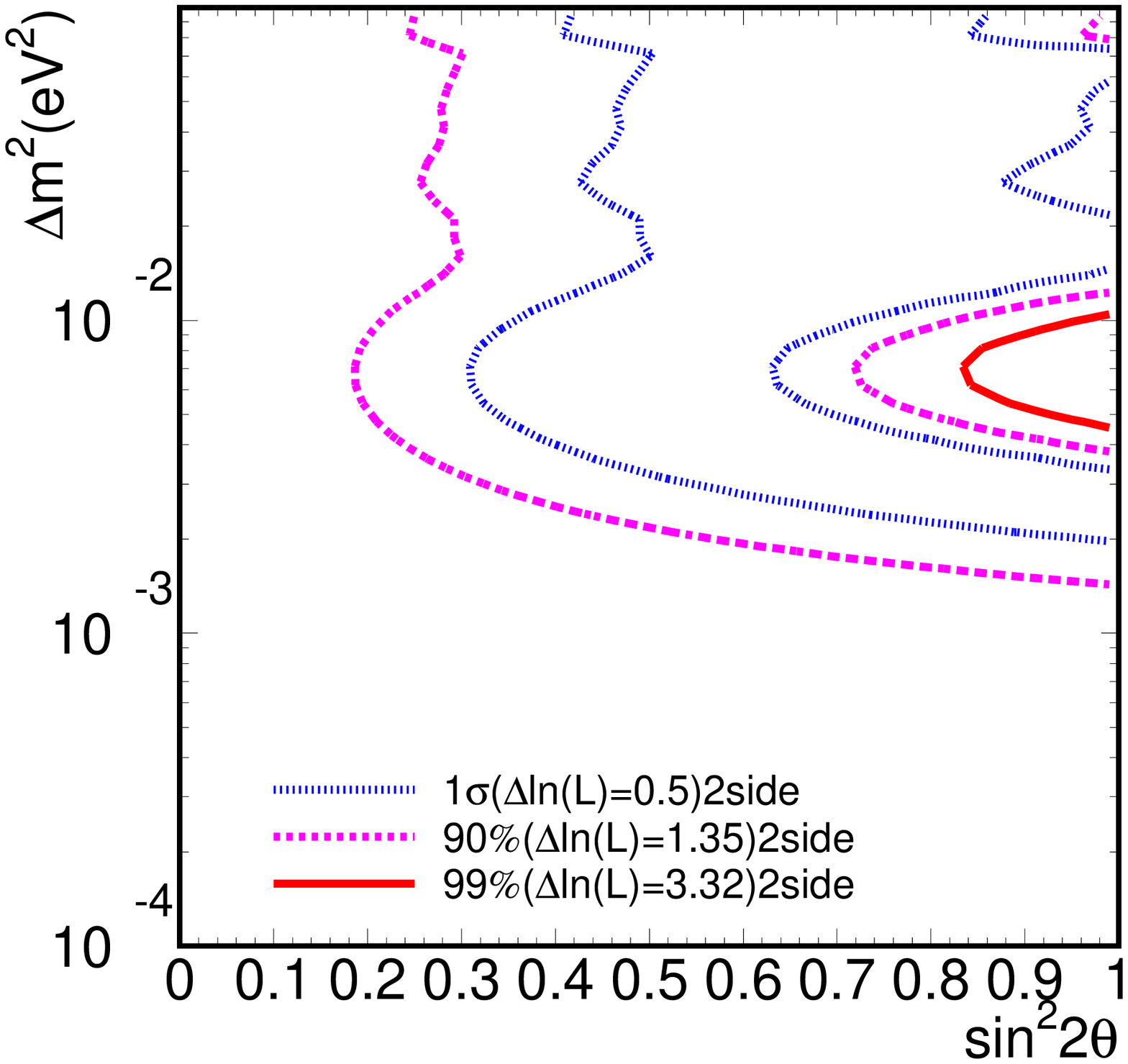}
\includegraphics[width=6cm]{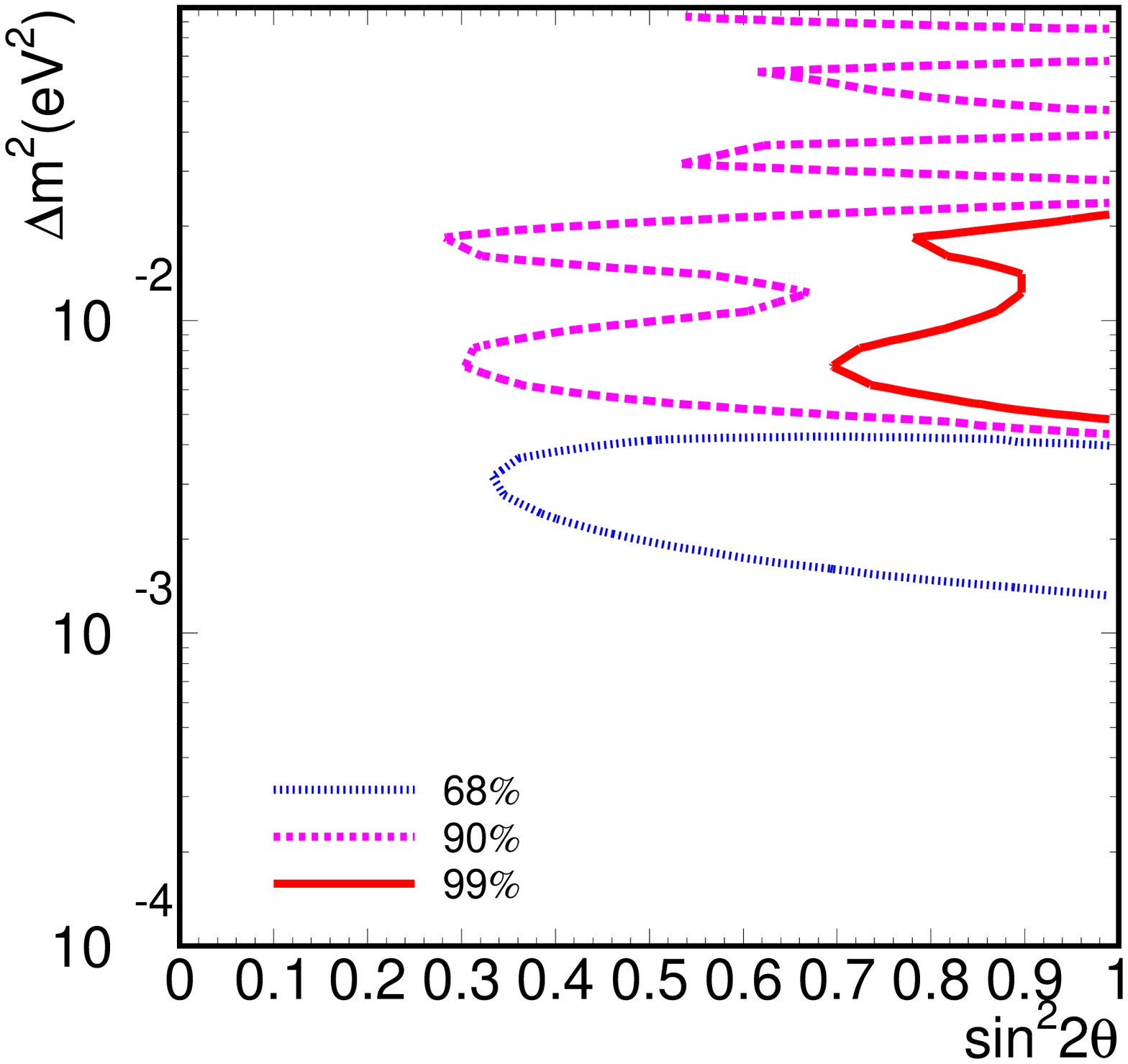}
 \caption{\it
The 68\%, 90\% and 99\% C.L. allowed region from K2K oscillation analysis.
The left contour is the result from the normalization analysis,and 
the right contour is from the shape analysis.
    \label{k2k7} }
\end{center}
\end{figure}

\begin{figure}[hbtp]
\begin{center}
\includegraphics[width=8cm]{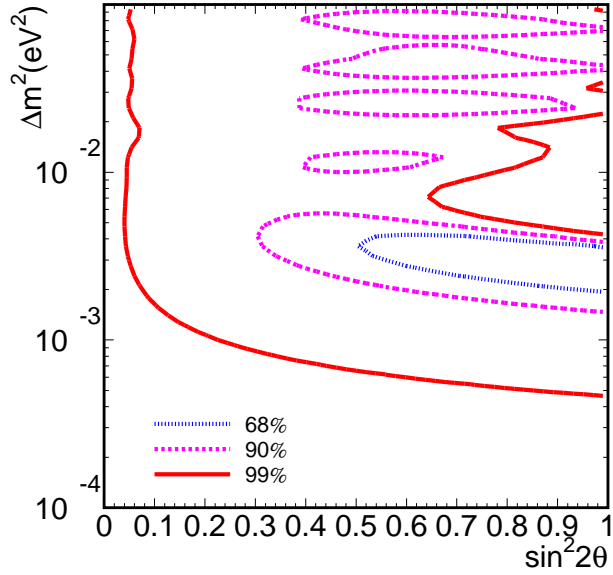}
 \caption{\it
The 68\%, 90\% and 99\% C.L. allowed region from K2K oscillation analysis.
    \label{k2k8} }
\end{center}
\end{figure}

\begin{figure}[hbtp]
\begin{center}
\includegraphics[width=6cm]{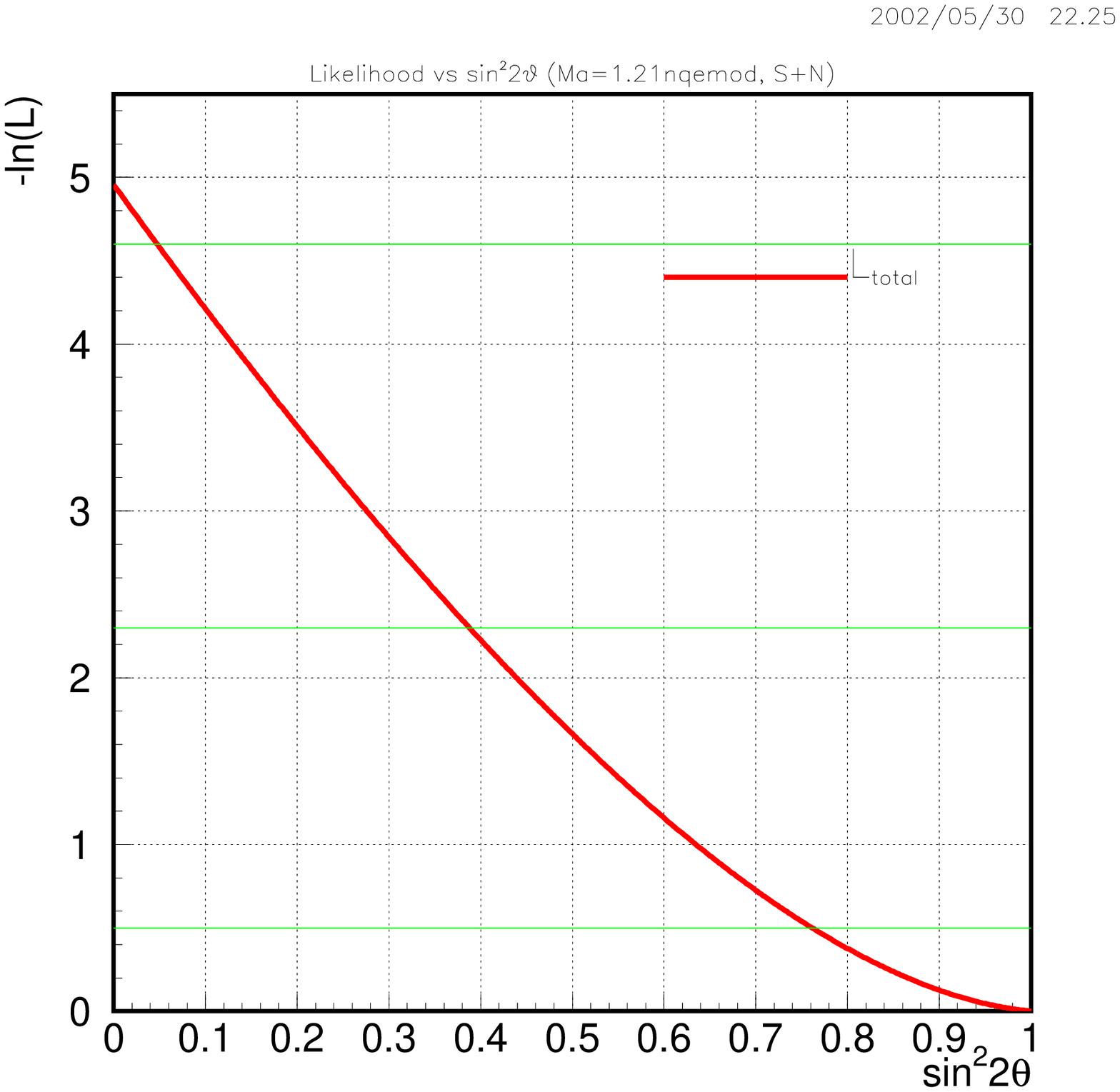}
\includegraphics[width=6cm]{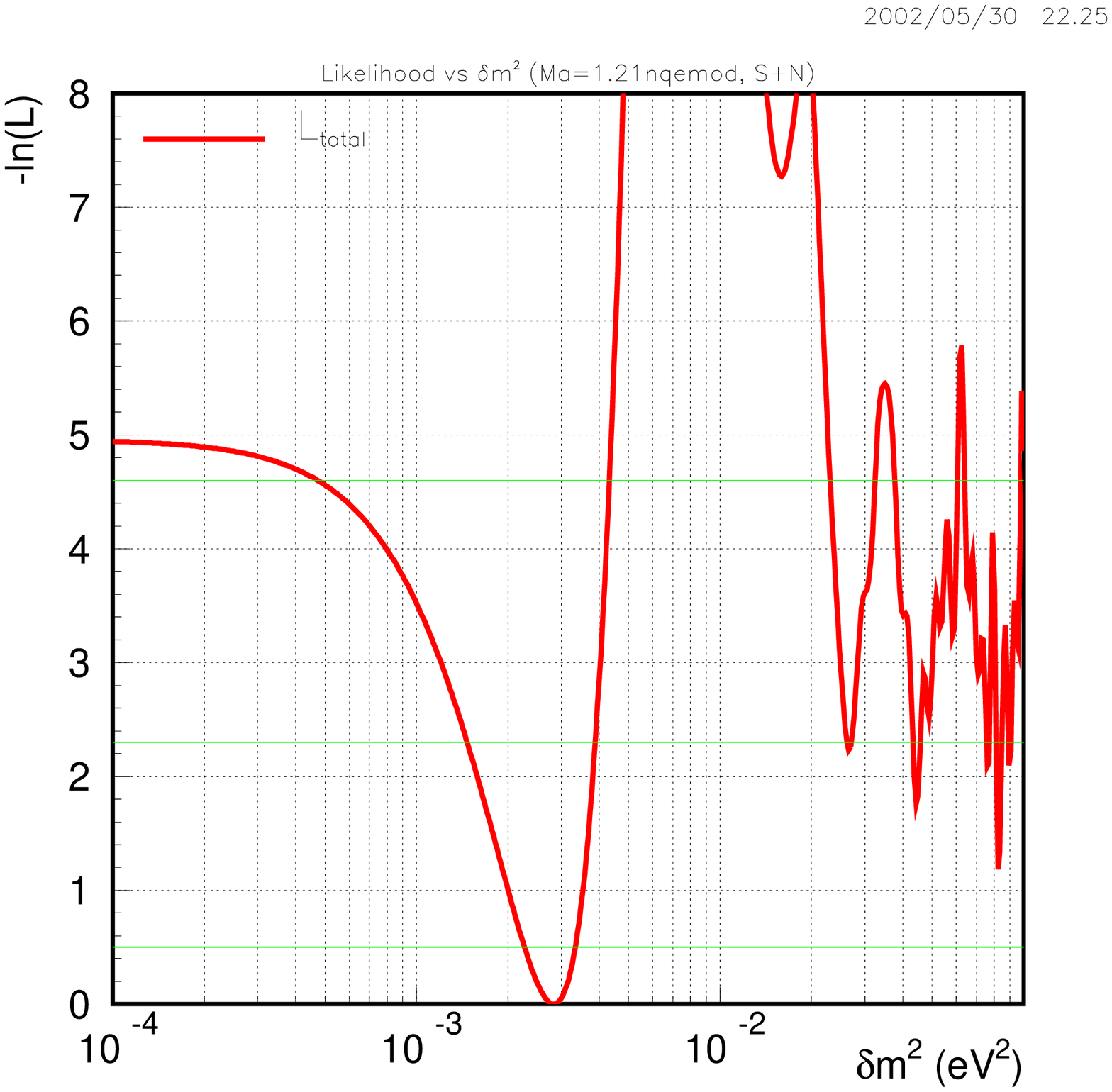}
 \caption{\it
The delta log likelihood as a function of $\sin^2 2\theta$ (left)
and $\Delta m^2$ (right).
    \label{k2k9} }
\end{center}
\end{figure}

The oscillation analysis is performed by the maximum-likelihood method. 
Both the number of observed events and the $E_\nu^{rec}$ distribution
is used in this analysis.
The likelihood is defined as 
\begin{equation}
{\cal L} = {\cal
  L}_{norm}(\Delta m^2, \sin^22\theta, f) \times {\cal
  L}_{shape}(\Delta m^2, \sin^22\theta, f) \times {\cal
  L}_{sys}(f). \label{llh}
\end{equation}
The normalization term is the Poisson probability to observe $N_{obs} (=56)$ 
when expected number of events is $N_{exp}(\Delta m^2, \sin^22\theta, f)$. 
The shape term is the product of the
probabilities of each one ring $\mu-like$ (1R$\mu$) event to be 
observed at $E_\nu^{rec}=E_i$;
\begin{equation}
{\cal L}_{shape}=\prod_{i=1}^{N_{1R\mu}} P(E_i; \Delta m^2, \sin^2 2 \theta, f),
\end{equation}
where $P$ is the normalized $E_\nu^{rec}$ distribution estimated by MC
simulation and $N_{1R\mu}$ is the number of 1R$\mu$ events.
The ${\cal L}_{sys}$ is the constraint term by systematic uncertainties.
The maximum likelihood fit is used in the analysis of two-flavor oscillation. 
The allowed region in 
$\Delta m^2$ and $\sin^2 2 \theta$ plane is shown in Figure~\ref{k2k7}
for normalization term and shape term separately.
Both the reduction and the energy distortion of muon neutrinos
indicate the neutrino oscillation with the same oscillation parameters.
The best fit point in combined analysis is 
$\rm \Delta m^2=2.8\times 10^{-3}$~$\rm eV^2$ and ${\rm sin^22\theta=1.0}$. 
The result is shown in Figure~\ref{k2k8}.
The probability
that the measurements at SK were explained by statistical
fluctuation without neutrino oscillation is less than 1\%.
In Figure~\ref{k2k8}, the allowed region of $\Delta m^2$ is
between $1.5\times 10^{-3}$ and $3.9\times 10^{-3}$~$\rm eV^2$ at
90\% C.L. at $\sin^2 2 \theta = 1.0$.
The delta log likelihood as a function of $\Delta m^2$ and $\sin^22\theta$
is shown in Figure~\ref{k2k9} which indicates the size of errors in this
analysis.

\newpage

\section{Summary}
\begin{figure}[hbtp]
\begin{center}
\includegraphics[width=12cm]{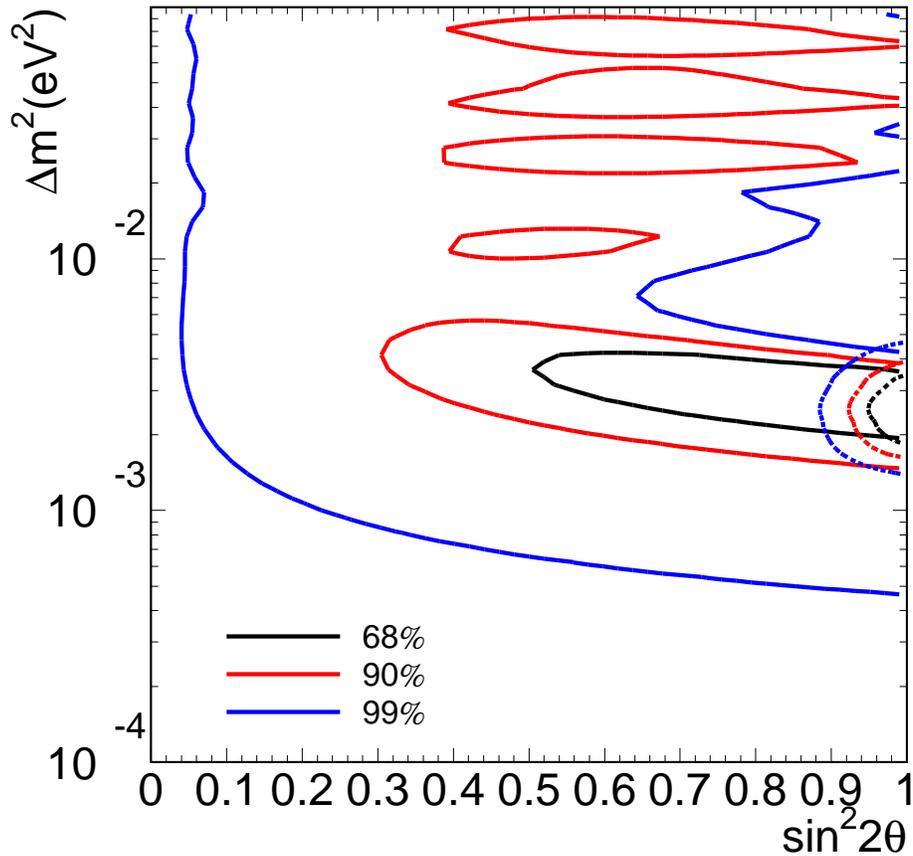}
 \caption{\it
The 68\%, 90\% and 99\% C.L. allowed region.
The SK atmospheric neutrino result (dashed line) and K2K
accelerator neutrino results (solid line) are overlaid.
    \label{sk2k} }
\end{center}
\end{figure}

Both SK atmospheric neutrino results and K2K accelerator neutrino 
results provide new information of neutrino oscillation.
The two results are shown in Figure~\ref{sk2k}, and 
agree well each other with $\Delta m^2$ of 
$(1.5-3.9)~\times 10^{-3}$~$\rm eV^2$ at 90\% C.L..
The neutrino oscillation is firmly established by SK,
and confirmed by K2K at $2.7 \sigma$ level.
The SK plans to resume the observation of atmospheric neutrinos 
around the end of 2002. 
The K2K plans to collect twice as many data as this by 2004.
Both experiments will provide more precise information of neutrino 
oscillation.

\end{document}